\documentclass[]{rmaa}

\usepackage{paralist}

\usepackage[latin1]{inputenc}

\usepackage{graphicx}

\title{Planetary
    nebulae in 2014:
     A review of research }

\author{Albert Zijlstra\altaffilmark{1} }
\altaffiltext{1}{ School of Physics and Astronomy, University of Manchester, UK.}

\shortauthor{Zijlstra}
\shorttitle{The PN review 2014}
\fulladdresses{
\item Albert A. Zijlstra: Jodrell Bank Centre for Astrophysics,  
  School of Physics and Astronomy,  University of Manchester,
  Oxford Road, Manchester M13 9PL, UK (a.zijlstra@manchester.ac.uk).}

\listofauthors{A. A. Zijlstra}
\indexauthor{Zijlstra, A. A.}

\SetYear{2015}
\ReceivedDate{20 May 2015}
\AcceptedDate{24 June 2015}
\SetVolume{51}

\abstract{ Planetary nebulae had a double anniversary in 2014, 250 years
since their discovery and 150 years since the correct spectroscopic
identification.  This paper gives an overview of planetary nebula
research published in 2014.  Topics include surveys, central
stars, abundances, morphologies, magnetic fields, stellar population
and galactic dynamics. An important continuing controversy is the
discrepancy between recombination-line and forbidden-line
abundances. A new controversy is the relation between symbiotic stars
and [WC] stars. PN of the year is undoubtedly CRL\,618, with papers on
its binary symbiotic/[WC] nucleus, rapid stellar evolution, expanding
jets and magnetic fields.}

\resumen{}

\addkeyword{Planetary nebulae}
\addkeyword{Stars: AGB and post-AGB}
\addkeyword{Stars: white dwarfs}
\addkeyword{ISM: abundances}
\addkeyword{ISM: jets and outflows}
\addkeyword{Galaxies: stellar content}

\begin{document}

\maketitle

\section{Introduction}

2014 was an important year for the study of planetary nebulae.  The first
planetary nebula (PN) was discovered on 12 July, 1764 when Charles Messier
stumbled across the Dumbbell Nebula, M27, making 2014 their 250th anniversary
year. They were quickly recognized as the most puzzling objects in the
sky. Antoine Darquier was the first to point out the similarity to the disk of
a planet. William Herschel concurred. He found several, including the Saturn
nebula, and wrote `{\it A curious nebula, or what else to call it I do not
  know\,}'. Herschel described them as seemingly a planet, but '{\it of the
  starry kind\,}'. The name `planetary nebula' captured the confusion well and
has been used ever since \citep{2014JHA....45..209H}. Coincidentally, 2014 is
also an important 150th anniversary: the first ever spectrum of a planetary
nebula (the Cat's eye nebula) was obtained by William Huggins on August 29,
1864, and it finally revealed their true nature 
  \citep{1864RSPT..154..437H}. Huggins wrote `{\it I looked into the
  spectroscope. No spectrum such as I expected! A single bright line only! At
  first I suspected some internal displacement of the prism. Then the true
  interpretation flashed upon me. The light of the nebula was
  monochromatic. The riddle of the nebulae was solved. The answer, which had
  come to us in the light itself, reads: not an aggregation of stars, but a
  luminous gas\,}' \citep[reported in][]{2007JBAA..117R.279M}. 

A double anniversary seems appropriate for objects of such dual nature.

Planetary nebulae still remain happily confusing objects, intruding into a
wide range of research areas. They trace stellar masses ranging from 1 to
$\sim 6$M$_\odot$.  Over 90\%\ of stellar death involves a PN phase, however
short lived. The very high luminosity, and the fact that up to 15\%\ of the
stellar luminosity can come out in a single emission line, makes PNe visible
out to very large distances, and makes them ideal kinematic tracers of dark
matter and of abundances of stellar populations, including areas where there
is little or no young stellar population. The morphologies point at the
physics of the ejection process, affected by rotation, binarity and magnetic
fields.

This article will summarize PN papers published in the refereed
literature in 2014, to identify the recent progress made in the field.
Over 100 journal papers are covered in this review. The review should
be read in conjunction with the White Paper produced by the IAU PN
Working Group, which identifies future research directions
\citep{2014RMxAA..50..203K}.


\section{Surveys and discoveries}

\subsection{The Galaxy}

IPHAS, a high resolution wide-area CCD H$\alpha$ survey of the
Northern Galactic plane, produced a catalog containing 159 new PNe
\citep{2014MNRAS.443.3388S}, with confidence in the classification
varying from `certain' to `possible'. Four further PNe were identified
from IPHAS by \citet{2014A&A...563A..63H}, { and one by
\citet{2014A&A...567A..49R} but the latter also re-classify one known PN
as a D-type symbiotic, leaving zero net gain.}  The corresponding CCD
survey of the southern galaxy, VPHAS, has commenced: the overview paper
\citep{2014MNRAS.440.2036D} shows the improvement over the photographic
SHS survey which has been responsible for many of the PN discoveries
in the south. The SHS was finally photometrically calibrated this year
\citep{2014MNRAS.440.1080F}, providing H$\alpha$ fluxes for 88 PNe.

\begin{figure*}
\includegraphics[width=15cm]{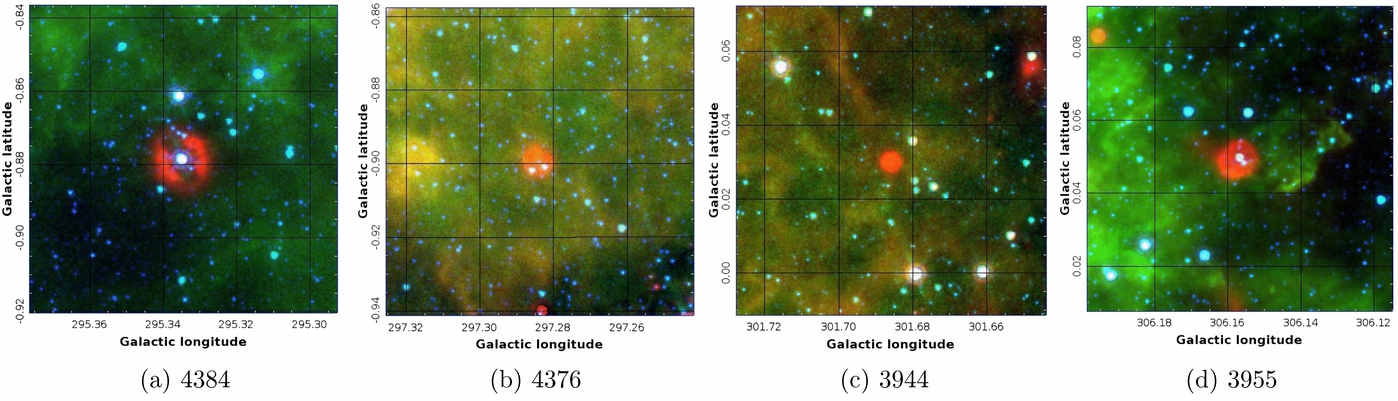} 
      \caption{Three-color images of representative Spitzer midgal bubbles.
        Red is MIPS 24$\mu$m, green is IRAC 8$\mu$m, and blue is IRAC
        4.5$\mu$m). Each image is 5' by 5'. From \citet{2014ApJ...796..116N},
        reproduced by permission. }
    \label{midgal}
\end{figure*}

The Spitzer galactic plane survey MIPSGAL has found many compact,
circumstellar bubbles emitting at 24$\mu$m. These bubbles (see
Fig. \ref{midgal} for examples) have attracted much attention. They include a
population of PNe, as shown by \citet{2014ApJ...796..116N} based on Spitzer/LRS
spectra, and by \citet{2014MNRAS.445.4507I, 2014MNRAS.437.3626I} 
based on JVLA radio observations. The suggestion that the bubbles have peculiar
central stars is an as-yet unproven speculation which bears investigating.

A novel survey is mapping the Milky Way in the shock-sensitive line of [Fe{\sc
    ii}] \citep{2014MNRAS.443.2650L}. Initial results detect 6 PNe out of 29
in this line. The second phase of the Chandra X-ray survey of nearby PNe was
published \citep{2014ApJ...794...99F}. The detection rates of diffuse X-ray
emission is 27\%\ and of X-ray point sources is 36\%. Diffuse emission is
mainly seen in young, compact PNe with closed shells, and is rare in bipolar
PNe.

One item of `It's a pity' research: the largest PN on the sky, Ou\,4
with a size exceeding 1 degree, may no longer be a PN. It seems to
originate from an H{\sc ii} region \citep{2014A&A...570A.105C}.

\subsection{And beyond}

Extra-galactically, the biggest advance, albeit negative, was from
PHAT (Panchromatic Hubble Andromeda Treasury; the acronym is far from
obvious). It re-classifies 32 nebulae in M31 as PNe, but on the other
side shows that 152 previously classified PNe are not
\citep{2014ApJ...792..121V}. No new objects were found, implying that
ground-based observations still are better at finding PNe but HST
is better at classifying them. There are now 591 PNe in the M31 PHAT
survey area (not all detected in PHAT), a change of $-17$\%. Up to 160
new PNe were found in the outer regions of M31 using SDSS data
\citep{2014AJ....147...16K}.

The Herschel Heritage project produced a catalog of far-infrared
sources in the Magellanic Clouds which contains a small number of
known PNe \citep{ 2014AJ....148..124S}. The total number of PNe in the
LMC is now 715 \citep{2014MNRAS.438.2642R}.


\section{Central stars}

\subsection{Binaries}
\label{bins} 

We lost one star, when the symbiotic star DT Ser was shown to be neither the
central star of the much more distant PN MSC1 at this position, nor a
symbiotic star \citep{2014MNRAS.445.1605F}. This loss was compensated by
discoveries of new binary companions elsewhere, so that 2014 was overall a
year of stellar increase, befitting a double anniversary.

Two long period binaries with periods of 1100 days and $>1800$ days were
discovered by \citet{2014A&A...563L..10V}, for PN G052.7+50.7 and LoTr\,5.  The
PN IPHASXJ211420.0+434136 (Ou\,5) gained a binary companion with a period of
8.4\,h \citep{2014MNRAS.441.2799C}, and Kn\,61 has a possible binary
companion, with a period of 5.7 days \citep{2014AJ....148...57G}: its nebula
is hydrogen deficient. NGC\,2392 may have a 0.12\,d companion, from radial
velocity variations \citep{2014MNRAS.440.2684P}; the star shows a strongly
variable wind.  A 0.61\,d binary system was uncovered in He\,2-11
\citep{2014A&A...562A..89J}: the post-common-envelope system is found to have
ended the AGB early.

NGC\,246 gained a star, when a tertiary companion to the binary system was
found (only the second triple system found among PNe). The two main-sequence
companions have masses of 0.85 and 0.1M$_\odot$ \citep{2014MNRAS.444.3459A}.
The central star is evolving erratically, possibly due to the recovery
from a symbiotic outburst 20 years ago. \citet{2014A&A...561A...8H} searched
for PN central stars binaries in the SMC and found 7 mimics (including one
symbiotic) and one true close binary, Jacoby SMC 1. They derive a
post-common-envelope fraction in the SMC of $<7$\%, below the 12--21\%\ of
the Galactic Bulge.

CRL\,618 gained a star \citep{2014ApJ...795...83B} when the central system was
shown to be a symbiotic-like system with a [WC8] companion, and the central
star of K\,3-18, previously suspected to be a symbiotic, was found to be a
[WC9] star \citep{2014MNRAS.440.1410M}.  The possible relation between
symbiotic stars and [WC] stars deserves further investigation, for those
wishing to complicate stellar evolution even further.

\subsection{Emission-line stars}

New stellar abundances were derived for five of the hottest [WC] stars
\citep{2014MNRAS.442.1379K}, showing a wide range in the C:He ratio,
high Ne abundances, and N abundances much higher than found in
previous determinations. 

Six new emission line stars were discovered by \citet{2014A&A...570A..26G},
including two [WC] stars and 1 VL (or [WC11]) star, making a total of 8 new
[WC] stars this year. They also discuss the issue of mimics among the
emission-line stars, where a nebular line can pretend to be a stellar one. You
have been warned.  The PG1159 (pulsating) stars are found to have the same
frequency of dust disks as other PN central stars \citep{2014AJ....147..142C}.

The mysterious O(He) central stars (H-poor, O-rich) were investigated by
\citet{2014A&A...566A.116R} who manage to both confirm and deny a merger
origin. A future in politics beckons. Refreshingly honest (which may argue
against their political future), they conclude that these stars exist and
therefore do form, but it is not clear how. \citet{2014MNRAS.440.1345F} argue
against a relation to the [WN] stars, and propose 'exotic channels', leaving
the evolution even more in darkness.

The small and equally mysterious group of [WN] stars increased its number to
four \citep{2014MNRAS.440.1345F} or five \citep{2014MNRAS.441.3065S} this year
(including one LMC PN), with the double re-classification of the star of A\,48
as [WN] \citep{2014MNRAS.441.3065S, 2014MNRAS.440.1345F}; the latter authors
list 6 further possible [WN] star candidates.  A\,48 either shows N/O in the
nebula four times above solar \citep{2014MNRAS.441.3065S}, three times above
\citep{2014MNRAS.439.3605D} or no significant N enrichment
\citep{2014MNRAS.440.1345F}, indicating either a higher-mass or lower-mass
progenitor star. The difference seems to be related to the derived oxygen
abundance. The nebula is highly carbon rich but the star has negligible carbon
\citep{2014MNRAS.439.3605D}. There seems some room for further research here.

Longmore 4 is an emission-line star which has varied between PG1159-type and
[WC].  The star has regular windy outbursts with high, hydrogen-poor and
helium-rich, mass loss, occuring approximately every 100 days
\citep{2014AJ....148...44B}. The paper finds 'no entirely satisfactory
explanation' and urges people to monitor similar stars.

\subsection{Normal stars}

These lack exoticism and are not as popular. The sole relevant paper of 2014
is on 63 hot post-AGB stars in the LMC which were found by
\citet{2014MNRAS.439.2211K}, as a byproduct of a survey, but they are not
known to have PNe around them.

\subsection{Magnetic fields}

Magnetic fields in the central stars remain elusive.  A sample of 12 stars
gave three possible, weak detections \citep{2014A&A...570A..88S}. The only
strong detection came from the A-type companion star to NGC\,1514.  Another
paper \citep{2014A&A...563A..43L} provided 23 non-detections, giving
2-$\sigma$ upper limits to the field strength of 500\,G to 6\,kG, whilst a
different permutation of these authors \citep{2014ApJ...787..111A} analysed 13
of the same targets with the same data but using a Bayesian approach to give
2-$\sigma$ upper limits of $<400$\,G. A targeted, high intensity observing
campaign may be needed to make progress in this area, but the evidence
indicates that observers should be experienced fly-fishers as not many stars
may bite. A solid detection would be a major advance, though.

\subsection{Evolutionary tracks}

The AGB and post-AGB is the most uncertain phase of the evolution of
single stars.  The standard Sch\"onberner tracks need to be accelerated by a
factor of 3, in order to fit central star masses
\citep{2014A&A...566A..48G}. This can be achieved by reducing the remaining
envelope mass at the end of the AGB: this is a free parameter in the mass-loss
models.

{ Evolution of stars up to the PN phase is reviewed by
\citet{2014PASA...31...30K}, with a particular focus on nucleosynthesis. The
agreement between stellar evolution models and PN abundances is generally
good, but enhanced oxygen in the intershell may be needed.}

\section{Morphologies}

\subsection{Origin}
The origin of the morphologies of PNe remains as elusive as ever. The dominant
shaping mechanism is magnetic fields, or binary motion, or stellar rotation
(but see below), or a combination of these; this leaves the details to be
worked out. There is progress with the growing evidence that the morphology is
determined on the AGB rather than the post-AGB, with detections of disks in
IRC +10$^{\rm o}$216 \citep{2014A&A...572A...3J} and L$_2$ Pup
\citep{2014A&A...564A..88K}.

\subsection{Cores, jets and lobes}
HST imaging of 10 compact PNe \citep{2014ApJ...787...25H} shows that
multipolar structures are common, indicating multiple phases of ejection and
shaping.  The paper does not state how the target sample was selected, so that
rates of occurence cannot be inferred. A new member of the small class of PNe
with extended shells and a compact, high density core was found
\citep{2014MNRAS.442..995M}, joining objects such as KjPn\,8, EGB\,6 and
M\,2-29. The core of Hb\,12 was studied by \citet{2014AJ....148...98C} using
infrared integral field spectroscopy, showing a system of bipolar lobes and
equatorial arcs. A compact infrared source with precessing lobes
\citep{2014A&A...561A..81B} was seen in K\,3-35, one of the few PNe with water
masers.  Spectroastrometry with CRIRES \citep{2014A&A...566A.133B} has
revealed structures as small as 12mas inside SwSt\,1 and IRAS\,17516$-$2525,
on par with VLTI resolution and much higher than achievable with HST.  The VLT
has since removed CRIRES for two years to allow for an upgrade, slightly
impeding this research. For now, high spatial resolution continues to be
dominated by HST. { \citet{2014A&A...569A..50C} present images and spectra
  for MyCn\,18: its asymmetries are attributed to a past explosive events,
  possibly an ILOT \citep{2012ApJ...746..100S} caused by planetary accretion.}

The role of magnetic fields received a boost with the finding of
well-ordered fields along the polar outflow axis in CRL\,618 and
OH\,231+04.1 \citep{2014MNRAS.438.1794S}, from dust polarization. 
Silicate dust shows higher polarization than carbon dust.

The expansion of the jets in CRL\,618 was re-analyzed from HST images
by \citet{2014A&A...561A.145R}: they find a Hubble-type flow.  The
mirror symmetry in the lobes of CRL\,618 may show indications of
alternate ejection in the two polar directions
\citep{2014ApJ...794..128V}, caused by one-sided heating of the
accretion disk. The term 'floppy disk' comes to mind (younger readers
may need to look this up). Discrete ejection events, albeit simultaneously
in both polar directions, during a common envelope phase were posited
by \citet{2014MNRAS.441.2799C}. 

{ For post-common-envelope PNe with jets, \citet{2014MNRAS.439.2014T} find
that in 3 out of 4 cases the jets formed a few thousand years before
the PN ejection occured, and in the 4th, more complex case, the jets
formed a few thousand years after the PN ejection. The latter case would
require very strong magnetic fields for the jet launching.}

Dusty disks around some PNe have been argued to be debris disks
\citep{2007ApJ...657L..41S}, or derived from binary-related post-AGB
gaseous disks \citep{2010A&A...514A..54G}. \citet{2014AJ....147..142C}
compare the two options and favour post-AGB disk although debris disks
cannot be ruled out.

\subsection{Models}

The highly complex structure of NGC\,6302 can be explained with an
isotropic wind blowing into a toroidal slow wind
\citep{2014MNRAS.442.3162U}. However, the observed Hubble wind requires
an additional acceleration, consistent with the observational result
of \citet{2011MNRAS.416..715S}. 

Jet models also remain
popular. \citet{2014MNRAS.440L..16B} investigate various accretion
models for driving jets from known pre-planetary nebulae, and rule out
most modes of accretion, including Bondi-Hoyle-Lyttleton wind
accretion and wind Roche-lobe overflow, based on observed jet power.
Roche-lobe overflow is possible, and accretion within a common
envelope could also explain observed jet momenta. The result depends
on the accretion time scale, which was reasonably chosen as the age of
the jet. { \citet{2014MNRAS.439.2014T} add magnetic fields to improve
accretion. They do not address the jet power problem, but their jets are older
(pre-dating the PN) and this may perhaps alleviate the power short-fall.}

Stellar rotation seems unable to
explain the shaping of bipolar PNe \citep{2014ApJ...783...74G}.

\subsection{Large scale structure}

At a time where much of the effort goes into studying small-scale
structures and jets, it is good to see that the overall large-scale
structure also is a rewarding research area.
{ \citet{2014AN....335..378S} study the expansion velocity structure of
PNe, based on 1-d radiation hydrodynamics. The paper gives a
much-needed description of the sometimes confusing terminology used:
`rim' for the inner-most shell compressed by the fast wind from the
star, `shell' for the dense region outside of the rim driven by
ionization pressure and surrounded by a shock, `double shell' if 
both are bright, and `halo' for the faint outer regions. The models
provide very good descriptions of the line profiles. The `rims' expand
at typically 10--20\,km\,s$^{-1}$, whilst the shells accelerate to up
to 50\,km\,s$^{-1}$. The acceleration makes it difficult to 
derive a kinematical age, or even a single expansion velocity (the
mass-averaged velocity can be useful at the 20\% level). The paper
also correctly stresses that the $v(r) \propto r$ Hubble wind, found
in bipolar flows and jets \citep{2014MNRAS.442.3162U,
  2011MNRAS.416..715S}, cannot be used to describe the main body of
the PN. }

Haloes around PNe are shielded from the UV field of the star by the dense
shell, and in 1-d models are often found to be recombining. But a detailed
study of the halo of NGC\,2348 finds that its halo, presumed to be
recombining, is in fact ionized and in equilibrium, with
$n_e=10$--30\,cm$^{-3}$ \citep{2014A&A...565A..87O}. The ionizing radiation
field may be leaking through the clumpy shell.

The interaction of the PN/AGB shell with the ISM is a continuing
avenue of research. \citet{2014A&A...570A.131V} show that the
interface can be shaped by the interstellar magnetic field into an
elongated feature, and when seen along the direction of the field can
show an `eye' shape. This paper is a candidate for the
title-of-the-year. Alignment of PNe, which remains a controversial
issue, has also been attributed to interstellar magnetic fields acting
on the ejecta \citep{2014MNRAS.438.2853F}, as opposed to acting on the other
end of stellar evolution \citep{2013MNRAS.435..975R}.

\section{Abundances}

PNe allow for easy but powerful abundance measurements. A good review
of PNe abundance studies and their controversies can be found in
\citet{2012IAUS..283..119K}.  It is an important area of study, with
impact on galaxy evolution, stellar dredge-up processes, and
indirectly, solid-state astrophysics.

\subsection{Theory and models}

The controversy regarding the mismatch between abundances
derived from forbidden lines and those from recombination lines
rumbles on, with proposed resolutions ranging from H-poor clumps to
non-Maxwellian electron velocity distributions \citep[the $\kappa$
  mechanism of][]{2012ApJ...752..148N}.  For oxygen, the recombination
lines are preferred by \citet{2014RMxAA..50..329P}.
\citet{2014MNRAS.440.2581S} strongly favour a two-temperature
distribution in Maxwellian equilibrium over a $\kappa$ non-thermal
distribution, but caution that their results may not apply to typical
PNe, whilst \citet{2014ApJ...780...93Z} cannot decide between these two
options. The war continues.

The excitation of the important forbidden 4363\AA\ line, used for
temperature measurements, was re-calculated by
\citet{2014MNRAS.441.3028S}. They find agreement with most older work
but disagreement with the recent result of
\citet{2012MNRAS.423L..35P}. New diagnostic diagrams and updated
formulae for electron temperature and density determinations are
presented by \citet{2014A&A...561A..10P}: they use the now
controversial calculations of \citet{2012MNRAS.423L..35P} for [O {\sc
    iii}] but re-introduce [Ar {\sc iii}]
(7135\AA\,+\,7751\AA)\,/\,5192\AA\ as an alternative
temperature-sensitive ratio at red wavelengths.

Abundance determinations from emission lines require correction for
unobserved ionization levels, through so-called ionization correction
factors. A large grid of these ICFs were calculated by
\citet{2014MNRAS.440..536D} using Rauch stellar models: the new grid
gives considerably different abundances for some elements, but also
provides a quantified indication of abundance uncertainties. This
could well become the highest cited paper of the year.

{ Oxygen is often assumed to trace the original abundances of the ISM
from which the star formed. However, third dredge-up does affect
oxygen abundances and can significantly increase its abundance at low
metallicity. Dredge-up is reviewed in \citet{2014PASA...31...30K}. }

\subsection{Practice}

An extreme nitrogen enhancement was found in NGC\,6302 by
\citet{2014A&A...563A..42R}. Herschel data for NGC\,6781 finds the
nebula to be mildly carbon rich, with C/O$\,=1.1$
\citep{2014A&A...565A..36U}.  Abundances were published for 53
Galactic PNe: these confirm the flattening of the [O/H] abundance
gradient in the inner Galaxy \citep{2014A&A...570A..26G}.

Iron is an important but difficult element. It is used for the
metallicity scale, but is strongly depleted even in ionized media due to
dust condensation. In PNe it is depleted by factors ranging from 2 to
500, with the highest depletions found in carbon-rich nebulae
\citep{2014ApJ...784..173D}.  Zinc provides an alternative
\citep{2014MNRAS.441.3161S}: it forms together with iron but is much
less refractory and doesn't suffer depletion. These authors find some
evidence for subsolar zinc abundances in Bulge PNe, with some
above-solar [O/Zn] values consistent with $\alpha$ element
enhancement.

\section{Molecules and dust}

\subsection{Masers and molecules}
The 5th case of an H$_2$O maser in a PN was discovered
\citep{2014MNRAS.444..217U}.  Water emission can be intermittent and
one post-AGB water-fountain star where the water maser disappeared,
presumed due to evolution towards the PN phase, caught astronomers out
by resurrecting its water maser \citep{2014A&A...569A..92V}. Who said
that evolution can't run backwards? The relation between water masers
in PNe and the 22 known water fountain sources remains to be
clarified. Methanol is still undetected in PNe
\citep{2014RMxAA..50..137G}, but the survey for it was limited to
oxygen-rich objects.

An impressive list of molecules was detected in M\,2-48, including
SiO, CO, CN and SO, as well as their isotopologues. The
$^{12}$C/$^{13}$C ratio of 3 is taken as evidence for hot bottom
burning \citep{2014ApJ...794L..27E} (however exchange reactions and selective
dissociation can alter the isotopologue ratios in partly ionized
regions). Molecular abundances in PNe show surprisingly little relation 
to the evolutionary age of the nebula \citep{2014ApJ...791...79E}, at
least in the five high mass bipolar nebulae which were observed. Molecules
seem to be resistant to dying.

One new molecule in PNe is OH$^+$, so good Herschel discovered it
twice \citep{2014A&A...566A..79A, 2014A&A...566A..78E}.

{ The field of molecules in PNe suffered a great loss with the passing
of Patrick Huggins. Unassuming but brilliant, always encouraging,
always ahead of his time with work on CO in evolved PNe, 
rings around AGB stars, depletion of refractory elements, and finally
jet lag in PNe: you never knew what would come next. We will miss him.}

\subsection{PAHs, fullerenes, and dust}
Some PNe, especially but not exclusively in the Galactic Bulge, show
both oxygen-rich dust and PAHs, indicative of mixed
oxygen-rich/carbon-rich chemistry. Images show that
the PAHs in these nebulae form at the outside of dense tori
\citep{2014MNRAS.441..364G}. The sample of Galactic PNe containing
fullerenes increased from 5 to 11 \citep{2014MNRAS.437.2577O} while
the number in the LMC remained 12 in 2014, thanks to a balance between the
removals and additions to the sample \citep{2014ApJ...791...28S}.

The PAH bands change during the PN phase \citep{2014MNRAS.439.1472M},
with an evolving mixture of aliphatic and aromatic features
\citep{2014ApJ...791...28S}, suggesting significant processing of the
molecules.  Dehydrogenation of PAHs by UV radiation could allow for
the formation of molecular hydrogen in PNe without requiring dust
\citep{2014MNRAS.441.1479C}.  Fullerenes are seen only in PNe with
cool central stars and apparently do not survive hard radiation fields
well \citep{2014ApJ...791...28S}.

Crystalline fosterite at 69\,$\mu$m was detected in five PNe using Herschel
spectroscopy \citep{2014A&A...565A.109B}. All have dense
tori, cool dust, and almost Mg-pure fosterite, i.e., very little iron
has entered the grains. Gas-to-dust ratios were measured in NGC 6781,
with values varying from 3000 in the centre to $\sim 200$ in the outer
regions \citep{2014A&A...565A..36U}. Sofia images show that the dust
in M\,2-9 is not confined to its disk but also spread through the
lobes \citep{2014ApJ...780..156W}.

\section{Distances}

Distances to Galactic PNe remain an issue.  Uncertainties in distance
relations can be slightly reduced by using probability density
functions rather than fitting linear relations
\citep{2014MNRAS.440.2026V}, but the scatter remains dominated by the
diversity in PN properties.  { The proposed surface brightness--radius
relation \citep{2008PhDT.......109F} may be able to reduce the
scatter.} In the sample of PNe with well-constrained individual
distances, \citet{2014A&A...567A...1M} published a new distance to the
open cluster Andrews-Lindsay\,1, which contains a PN. The PN He\,2-86
may be a member of the cluster NGC\,4463 but this requires that the
extinction to the PN is partly internal \citep{2014A&A...561A.119M}.
Accurate distances for { individual} PNe are awaiting GAIA.

\section{Evolution}

PNe expand at typically 20--40 km/s: over decades this gives a
detectable motion. HST has been used to detect this expansion in the
hydrogen poor gas ejected by a Very Late Thermal Pulse (VLTP) in A30
and A78 \citep{2014ApJ...797..100F}.  The expansion does not show the
Hubble flow seen in bipolar PNe { (but note the discussion on PN
velocity fields in \citet{2014AN....335..378S})} but reveals a more
complex pattern of local acceleration and deceleration. Sakurai's
Object continues on its fast-track post-VLTP
evolution. \citet{2014ApJ...785..146H} have resolved the ejecta into a
fragmented outer region expanding at $10^3$\,km\,s$^{-1}$ seen in
shocked He{\sc i}, and a bipolar dust debris cloud expanding at
55\,$\mu$arcsec\,d$^{-1}$. { As a by-product, this introduces a new unit
into the field of PNe.}

The slow increase of central star temperature in PNe should cause an increase
in excitation, and in particular the strength of the [O{\sc iii}] lines. A
tendency for such an increase was found in NGC\,6572 over 80 years of
observations \citep{2014ARep...58..702A}, and in Hen 2-260 over 28 years of
data \citep{2014A&A...567A..15H}. In the latter object, the stellar
temperature is increasing by 45\,K\,yr$^{-1}$.  Three young PNe were found to
show fading central stars, attributed to increasing temperatures since the
epoch of the BD and CD catalogues, a century ago \citep{2014AstL...40..485A}.

\begin{figure*}
\includegraphics[width=15cm]{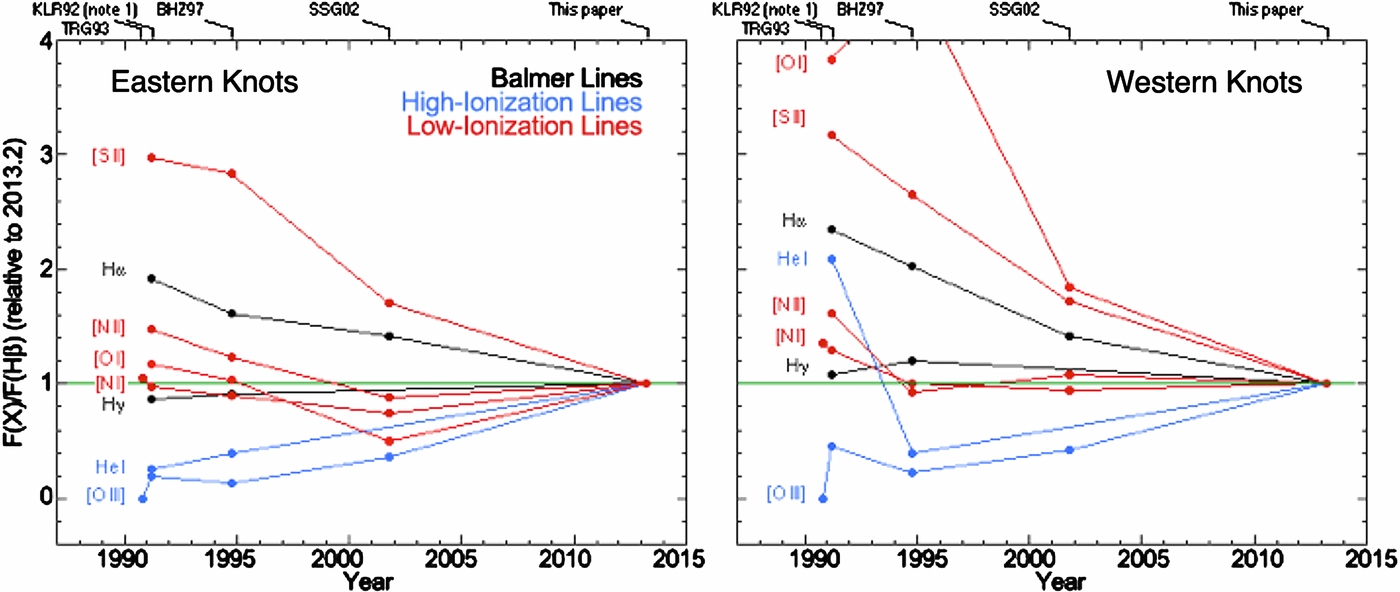} 
      \caption{Changes in line flux ratios of CRL\,618 relative to the data
        obtained during 2013.2.  Some of the scatter is measurement related.
        From \citet{2014ApJ...795...83B}, reproduced by permission. }
    \label{c618}
\end{figure*}

CRL\,618 shows evidence for an increasing central star temperature
\citep{2014ApJ...795...83B}, from its increasing radio flux and evolving line
ratios (see Fig. \ref{c618}). But the increase over 40\,yr is deemed to be
unreasonably fast, and UV changes due to symbiotic accretion is also
considered a possibility. This could move the object to Section \ref{rel}.

The hot bubble inside PNe should leak between the clumps of the PN shell and
depressurise. The process that stops this from happening seems to be mixing of
shell (or rim) material into the hot bubble \citep{2014MNRAS.443.3486T}.

\section{Stellar populations and extra-galactic PN}

\subsection{The Galactic bulge}

PN abundances in the Galactic bulge indicate progenitor stars of 3-5 M$_\odot$
\citep{2014A&A...567A..12G}. Such a young population would be surprising in
the Bulge, although \citet{2014A&A...566A..48G} also find evidence that the
PNe in the Bulge derive from a younger stellar population.

\subsection{Nearby galaxies}
PN oxygen abundances show a mass-metallicity relation of Local Group galaxies
\citep{2014MNRAS.444.1705G} in agreement with other abundances, except for the
lowest metallicity galaxy, Leo A, where the PN has more oxygen than
expected. The authors do not discuss whether oxygen dredge-up
\citep{2014PASA...31...30K} may have an effect on their relation at low
metallicity.

PNe in M81 and other spiral galaxies show a shallower oxygen abundance
gradient than do the H{\sc ii} regions \citep{2014A&A...567A..88S}. This
indicates the effect of pre-enriched infall into the galaxies, but stellar
migration may also play a role.  The PNe in NGC\,6822 trace the dynamics of
the intermediate-age stellar population, and not that of the H{\sc i} disk and
its young stellar population \citep{2014A&A...568A..82F}.

In M31, PNe in the outer regions  trace known morphological  features,
including the Northern Spur, the NGC 205 Loop, the G1 Clump, and And
NE, and indicate that the Giant Stream, the Northern Spur, and possibly
And NE are kinematically connected \citep{2014AJ....147...16K}. The
mass of the Giant Stream progenitor is estimated at $10^9$\,M$_\odot$
from its PNe.

\citet{2014MNRAS.442.3544F} have found 32 PNe in the Umbrella galaxy,
NGC\,4651, of which 10 are in the nearby stream tracing a disrupted
dwarf galaxy. They show that the dwarf was on a short-period orbit and
had recently passed through the disk of NGC\,4651. PNe thus trace not only
the death of stars, but the death of galaxies as well.

\section{Related objects}
\label{rel}

R Cor Bor stars do not produce PNe. However, they do have winds similar to
those briefly shown by VLTP objects such as Sakurai's Object.
\citet{2014A&A...569L...4C} show that the R Cor Bor star V854 Cen is producing
a mildly flattened shell which they term `bipolar'.  (A PN with that
elongation would have been `elliptical'.) The paper was written in hospital
and appeared four months after the untimely death of the first author, Olivier
Chesneau. He is sorely missed. We can be proud of his achievements, and take
consolation in the fact that the field of PNe attracts people as brilliant and
as universally liked as Olivier.

Overly rapid and erratic evolution can be a sign of a PN mimic. The central
star of the beautiful Stingray nebula was found to have increased in
temperature from 38\,kK to 60\,kK over 14 years, with a declining luminosity
(and radius), followed by a temperature decrease to 50kK over the next 4
years. The star is classified as sdO, but it could be a post-RGB star or/and a
product of common envelope evolution \citep{2014A&A...565A..40R}.

For the growing relation between symbiotic stars and PNe with [WC] central
stars I refer to Section \ref{bins}.

\section{Miscellaneous}

Planetary nebulae can make useful calibration sources, as they are compact and
bright over a very wide range of wavelengths. As an example, NGC\,7027 is an
important calibrator at radio wavelengths \citep{2013ApJS..204...19P}. The W4
filter of the Wise mission was re-calibrated using PNe 
\citep{2014PASA...31...49B}. This made use of the strong [O{\sc IV}] emission 
line at 25.9\,$\mu$m which dominates the wavelength range of the W4 filter.

Numerical simulations often require setting an initial distribution of
parameters over a grid. \citet{2014LNCS.8632..656H} present a method to create
a `random field' using GPUs, and apply it to defining the initial conditions
for an inhomogenous, turbulent wind of a PN. Sadly this is where they stop.

Physics in the strong gravity of degenerate stars was shown to be well-behaved
by \citet{2014PhRvL.113l3002B}: specifically, there is no evidence that the
proton-electron mass ratio depends on gravity.  PNe may be as confusing in
their own right as they were 250 years ago, but the laws of physics at least
apply.  But it remains a physics of complexity, bordering on physics of
perplexity.



\bibliography{pn2015_rmaa}

\end{document}